\documentclass[aps,prb,reprint]{revtex4-1}

\usepackage{amsmath,amsfonts,amssymb,graphicx,braket}
\usepackage[]{units}
\usepackage{hyperref}
\hypersetup{
    colorlinks=true,
    linkcolor=blue,
    filecolor=blue,      
    urlcolor=blue,
    citecolor=blue
}

\begin{document}
\title{Theory of Strain-Induced Confinement in Transition Metal Dichalcogenide Monolayers}
\author{Matthew Brooks}
\email{matthew.brooks@uni-konstanz.de}
\author{Guido Burkard}
\affiliation{Department of Physics, University of Konstanz, D-78464, Germany}

\begin{abstract}
Recent experimental studies of out-of-plane straining geometries of transition metal dichalchogenide (TMD) monolayers have demonstrated sufficient band gap renormalisation for device application such as single photon emitters. Here, a simple continuum-mechanical plate-theory approach is used to estimate the topography of TMD monolayers layered atop nanopillar arrays. From such geometries, the induced conduction band potential and band gap renormalisation is given, demonstrating a curvature of the potential that is independent of the height of the deforming nanopillar. Additionally, with a semi-classical WKB approximation, the expected escape rate of electrons in the strain potential may be calculated as a function of the height of the deforming nanopillar. This approach is in accordance with experiment, supporting recent findings suggesting that increasing nanopillar height decreases the linewidth of the single photon emitters observed at the tip of the pillar, and predicting the shift in photon energy with nanopillar height for systems with consistent topography.
\end{abstract}
\maketitle

\section{Introduction}
 \label{sec:Intro}
 
Transition metal dichalchogenide (TMD) monolayers are atomically thin semiconducting crystals boasting optically active, direct band gaps and strong spin-orbit coupling which in turn introduces optically addressable spin-valley coupling\cite{wang2012electronics,kumar2012electronic,xiao2012coupled,kormanyos2014spin}. Chemically, semiconducting TMD monolayers are described as $MX_2$, consisting of one transition metal atom $M=$ Mo or W for every two chalcogen atoms $X=$ S or Se arranged in a staggered hexagonal 2D lattice, similar to graphene but with a broken inversion symmetry. This allows for the electrons to possess the time reversal symmetric, valley-isospin degree of freedom ($K/K'$), while the broken inversion symmetry opens a direct, optical range band gap about these valleys. Additionally, the transition metal atoms introduce a strong spin-orbit coupling, correlating the spin and valley degrees of freedom, forming twofold-degenerate Kramers pairs $|K\uparrow\rangle/|K'\downarrow\rangle$ and $|K'\uparrow\rangle/|K\downarrow\rangle$. Since monolayer isolation, a number of possible devices exploiting the novel spin-valley and 2D material physics have been theorised and demonstrated. These include low dimensional Van der Waals heterostructure field effect transistors\cite{chuang2014high,jo2014mono,wachter2017microprocessor}, photovoltaic systems and photo-detectors\cite{tsai2014monolayer,feng2012strain,lemme2011gate} as well as spintronic\cite{zibouche2014transition,gmitra2016trivial,qian2014quantum} and valleytronic\cite{gong2012magnetoelectric,mak2012control,ye2016electrical} devices. 

Several standard material manipulation and combination techniques have already become part of the standard toolbox of monolayer engineering, including metal electrode gating and layered heterostructure composites\cite{chuang2014high,jo2014mono,wachter2017microprocessor,tsai2014monolayer,lemme2011gate,zibouche2014transition,gmitra2016trivial,qian2014quantum,gong2012magnetoelectric,mak2012control,ye2016electrical}. Recently, a number of studies into out-of-plane straining as a novel manipulation technique have been experimentally investigated for deterministically implementing quantum light sources\cite{palacios2017large,branny2017deterministic,kern2016nanoscale}. Similarly to other low dimensional crystals such as graphene and hexagonal boron nitride (h-BN), TMDs exhibit great flexibility and mechanical strength. It is known that TMD monolayers can withstand tensile strain up to the order of $10\%$\cite{coleman2011two} before rupture, thus the ongoing interest in TMDs for flexible substrate technologies\cite{velusamy2015flexible,gong2016high}. As such, there have been notable DFT studies into the electronic response of TMD monolayers to tensile strain\cite{chang2013orbital,guzman2014role}. Interestingly, it is believed that all TMD species form a type II quantum well (electron confining but hole repulsive) of the conduction and valence bands under strain, with the exception of WSe$_2$ which forms type I quantum dots (electron and hole confining). One noticeable change in the behaviour of specifically sulphide semiconducting TMDs (\textit{M}S$_2$) under strain is a direct to indirect band gap transition that has been observed at $2.5\%$ tensile strain in WS$_2$\cite{wang2015strain} and calculated to be at about $2\%$ for MoS$_2$\cite{chang2013orbital}.

Quantum light sources had previously been observed in a TMD monolayer at strained defect points along the edge of a monolayer flake\cite{koperski2015single}. With out-of-plane straining, this effect has now been shown to be deterministically implementable, by creating strain fields with an appropriate renormalisation of the band gap to funnel excitons to a given location in WSe$_2$ by placing the TMD on a substrate that selectively deforms the monolayer. Similarly, it has previously been suggested that an atomic force microscope (AFM) tip may be used to strain MoS$_2$ monolayers for electron collection in photovoltaic devices\cite{feng2012strain}. It is clear that the flexibility, strain band-response and durability of TMD monolayers opens up the novel device implementation tool of strain manipulation, by exploiting the third dimension of a 2D material. Out-of-plane strain field engineering has the potential to become part of the standard toolbox of TMD device implementation, to be used as an additional tool to help manipulate the conduction and valence bands. The potential for strain engineering for quantum emitters is now well demonstrated, but a similar method could be combined with other known manipulation techniques to allow for hybrid strain-gated electronic devices.

It is the goal of this work to develope a satisfiying approximate analytical model of the TMD topography due to a deforming element such as a nanopillar grown from a substrate. Thereafter the strain induced potentials from such geometries will be calculated and predictions of the bandgap renormalisation, single particle energy spectra and probability of tunneling out of the strain defined potential region will be made.

This Paper is structured as follows. An analytical description of a TMD monolayer deformed about a nanopillar grown out of a silica substrate is theoretically derived in Sec.~\ref{sec:Deformation} followed by an analysis of the strain-induced potential from the derived deformation in Sec.~\ref{sec:StrainPot}. In Sec.~\ref{sec:FckDrwEn} the energy levels of electrons confined by the strain induced potential are given and in Sec.~\ref{sec:WKB} the semi-classical WKB approximation is used to estimate electron leakage from the given potentials. Finally, a discussion of the possible devices single particle strain-induced potential wells allow is provided in Sec.~\ref{sec:Discussion} and a summary of the presented work is given in Sec.~\ref{sec:Summary}.

\section{Deformation Topography}
 \label{sec:Deformation}
 
In this work we calculate the out-of-plane deformation topography of the TMD monolayers layered atop nanopillars using continuum-mechanical plate-theory. The full set of elastostatic equilibrium equations\cite{landau1986theory} assuming rotational symmetry are
  
    \begin{subequations}
		\begin{equation}
			D\Delta^2\zeta-\frac{h}{r}\left(\frac{d \chi}{d r}\frac{d^2 \zeta}{d r^2}+\frac{d^2 \chi}{d r^2}\frac{d \zeta}{d r}\right)=P
			\label{eq:PlateEQ2a}
		\end{equation}
		\begin{equation}
			\Delta^2\chi+\frac{E}{r}\left(\frac{d\zeta}{dr}\frac{d^2\zeta}{dr^2}\right)=0
			\label{eq:PlateEQ2b}
		\end{equation}
    \end{subequations}

\noindent where $\zeta$ is the deformation coordinate (height field) of TMD, $\chi$ the stress function, $h$ the thickness of the TMD, $E$ is the Young's modulus, $P$ is the externally applied force per unit area and $D$ is the flexural rigidity of the TMD defined as

    \begin{equation}
        D=\frac{Eh^3}{12(1-\sigma^2)}
        \label{eq:flexuralRigidity}
    \end{equation}

\noindent where and $\sigma$ is the Poisson's ratio. The stress function $\chi$ is defined as

    \begin{equation}
        \Delta\chi=\frac{E}{(1-\sigma)}\nabla\cdot \mathbf{u}=\frac{E}{(1-\sigma)}\left(\frac{1}{r}\frac{d (r u_r)}{d r}\right)
        \label{eq:stressfunction}
    \end{equation}
	
\noindent where $\mathbf{u}=(u_r,u_\theta)$ is the displacement vector. In this work, we study similar topographies to those used in experiments where the TMD monolayer is passively strained (lacking clamping of the edges and allowing for elastic equilibrium). In such a regime, the contribution from the stress function to the overall strain in the TMD will be five oders of magnitude lower than that of the contribtion from the height field topography. As such, a pure bending regime is assumed where the elastostatic equations may be simplified to 

	 \begin{figure}[!ht]
	    \includegraphics[width=0.96\linewidth]{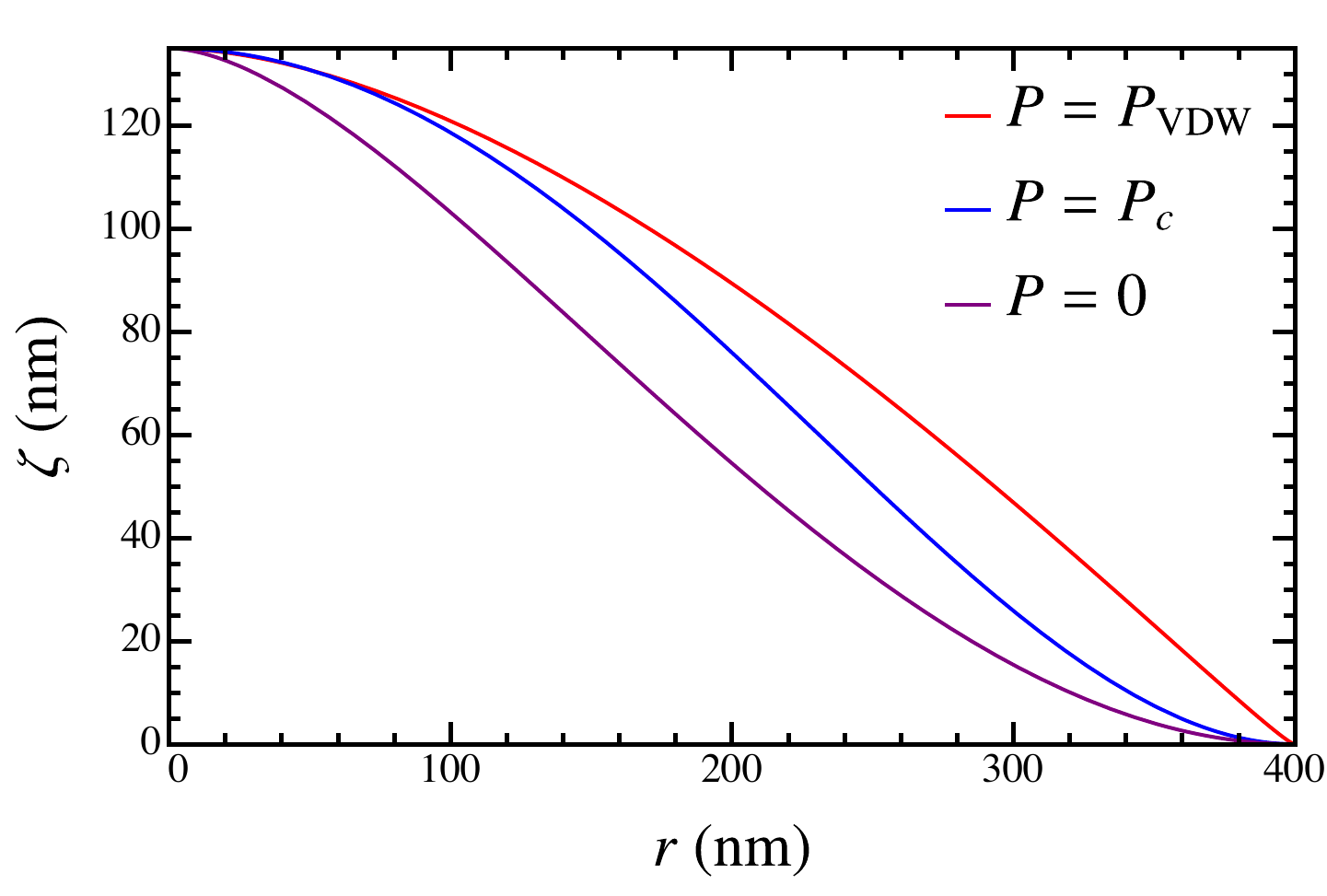}
	    \caption{2D radial deformation topography of WSe$_{2}$ derived from the $P=0$ (purple) $P=P_c$ (blue) and $P=P_{\textrm{VDW}}$ (Red) assuming a nanopillar height of $\unit[135]{nm}$ with tenting radius of $\unit[400]{nm}$ as chosen from experimental examples\cite{palacios2017large}.}
	    \label{fig:zeta_Models}
	\end{figure}

    \begin{equation}
        D\Delta^2\zeta-P=0
        \label{eq:PlateEQ}
    \end{equation}

\noindent where (\ref{eq:PlateEQ2b}) may no longer be satisfied. 

There are a number of choices for force per unit area to be considered; $P=0$ with boundary conditions, Van der Waals attraction between the TMD and substrate\cite{klimchitskaya2000casimir}

    \begin{equation}
        P_{\textrm{VDW}}=\frac{\mathcal{H}_{\textrm{TMD-Sub}}}{(h/2+\zeta)^3}
        \label{eq:VDWatt}
    \end{equation}

\noindent where $\mathcal{H}_{\textrm{TMD-Sub}}$ is the Hamaker constant between the choice TMD and substrate, and a constant force per unit area $P=P_c$ approximation. The Van der Waals force topography may be calculated numerically while the height fields of the $P=0$ and $P=P_c$ models may be exactly solved to give 

    \begin{equation}
		\zeta_{P=0}(r)=\frac{H\left(R^2+r^2\left[\log\left(\frac{r^2}{R^2}\right)-1\right]\right)		}{R^2}
        \label{eq:0PZeta}
    \end{equation}

\noindent and

    \begin{equation}
        \zeta_{P=P_c}(r)=\begin{cases}
    	\beta\left(\sqrt{\frac{H}{\beta}}-r^2\right)^2   & \quad r\leq R\\
   	 	0  & \quad r> R\\
  		\end{cases}
        \label{eq:RealSols}
    \end{equation}
	
\noindent respectively where $H$ is the height of the deforming nanopillar i.e. the height at which the TMD is held at at the origin, $R$ is the tenting radius i.e. the radius at which the TMD meets the substrate and $\beta$ is defined as

    \begin{equation}
        \beta=\frac{P_c}{D64}=\frac{3 P_c (1-\sigma^2)}{16 h^3 E}.
         \label{eq:betaDef}
    \end{equation} 

\noindent Both of these models for the height field assume ``clamped'' boundary conditions $\partial_r \zeta(r)\rvert_{r=0,R}=0$. The values of $R$ and $H$ need to be assumed for the $P=0$ model, as all mechanical properties of the TMD are lost, while for the $P=P_c$ model the relationship between $R$ and $H$ is given as $R=\sqrt[4]{H/\beta}$.
	
	 \begin{figure}[!t]
	    \includegraphics[width=0.96\linewidth]{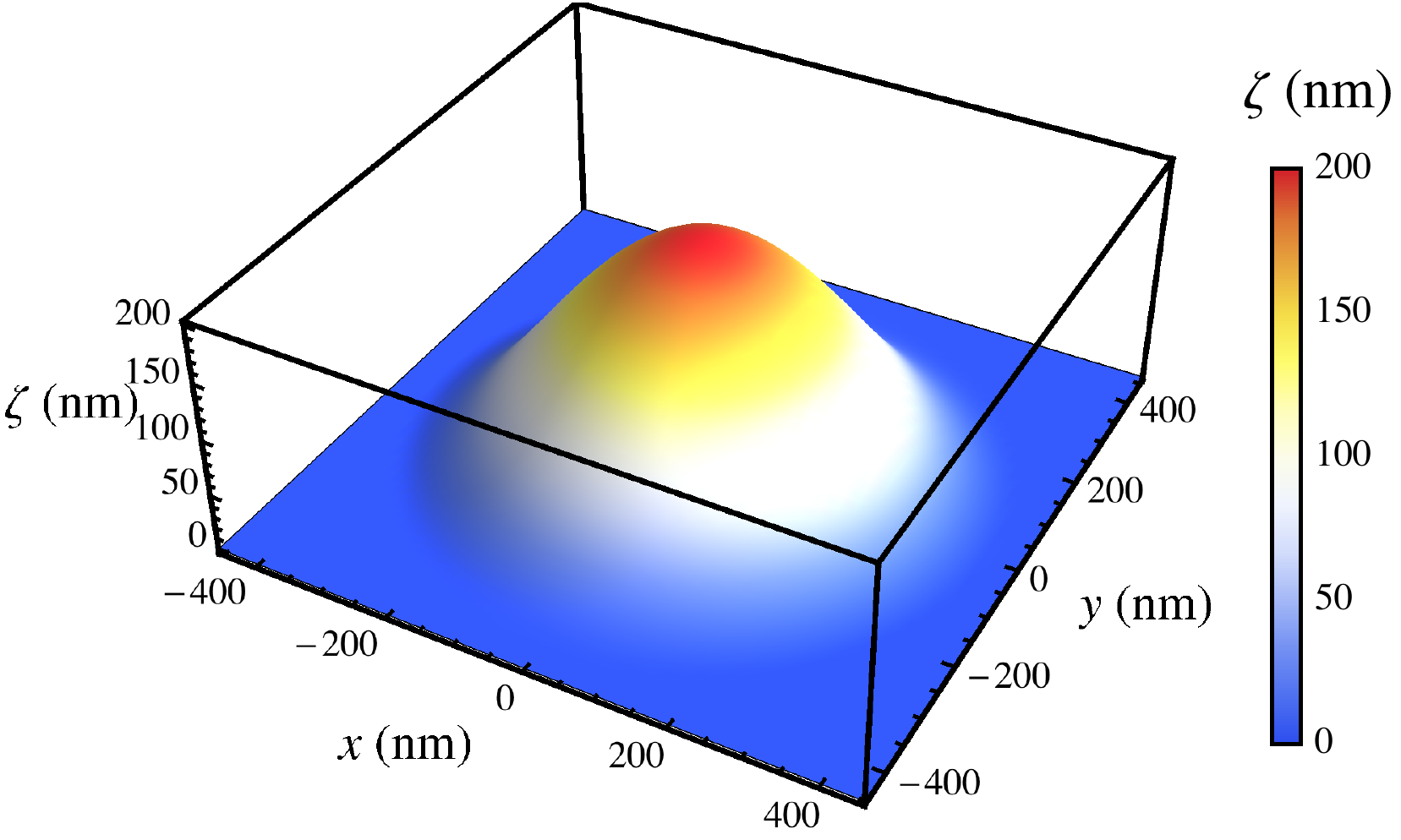}
	    \caption{Height field $\zeta(x,y)$ as a function of the $x$ and $y$ coordinates in the TMD plane of a WSe$_{2}$ monolayer layered atop a  $\unit[200]{nm}$ tall nanopillar.}
	    \label{fig:WSe2_3d_Topography}
	\end{figure}
	
Fig.~\ref{fig:zeta_Models} demonstrates the difference between the topographies given by the three proposed $P$ functions. As is evident, the $P=P_c$ model aligns well with the $P=P_{\textrm{VDW}}$ model close to the origin (where electron confinement takes place) while close to the tenting radius the $P=P_c$ model aligns with the $P=0$ model. As such, the $P=P_c$ will be used to give a reasonable analytical approximation to experimental works which we aim to model.
    
This work will focus on the TMD monolayer species of MoS$_{2}$ and WSe$_{2}$. WSe$_{2}$ is considered since it has been the focus of past TMD strain experiments\cite{palacios2017large,kern2016nanoscale,branny2017deterministic} that measured quantum emitters in strained regions of the monolayer. WSe$_{2}$'s optical response on and off resonance may be greatly enhanced\cite{wang2015giant} and, as has been shown in DFT studies\cite{chang2013orbital}, exhibits exciton funnelling under strain. MoS$_{2}$ is also considered, as this material has been studied for its possible spintronic and valleytronic applications such as quantum dots\cite{pisoni2017gate} for quantum information\cite{kormanyos2014spin,brooks2017spin} due to its relatively small spin-orbit splittings. Values for the Young's modulus\cite{bertolazzi2011stretching,morell2016high}, Poisson's ratio\cite{liu2014elastic,ccakir2014mechanical} and layer thickness\cite{splendiani2010emerging,huang2016probing} are all taken from mechanical experiements, while a reasonable value for the applied force is approximated from the tenting radii of a nanopillar strained TMD experimental study\cite{palacios2017large}.

\begin{figure}[!ht]
	    \includegraphics[width=\linewidth]{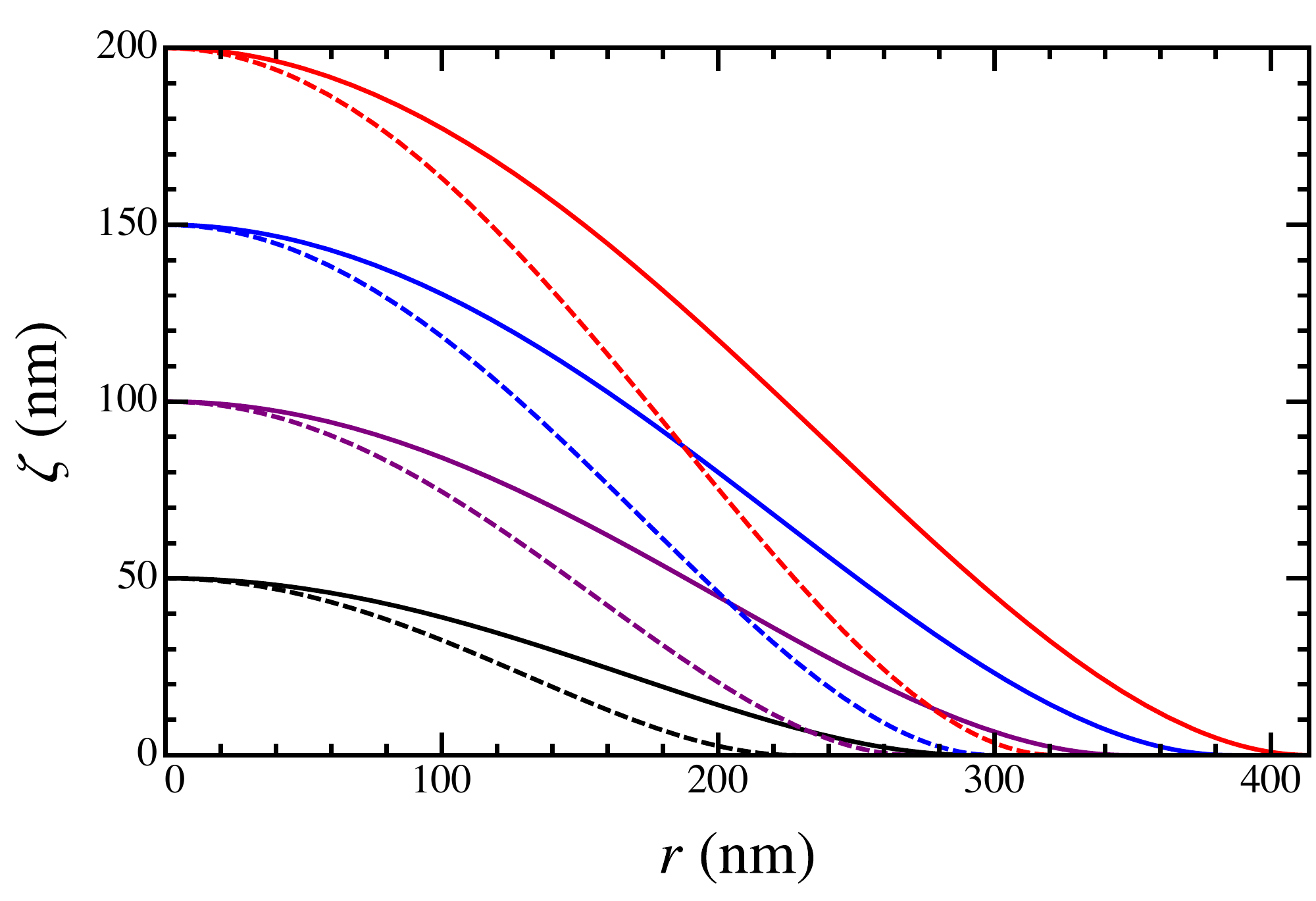}
	    \caption{2D radial deformation topography of WSe$_{2}$ (solid) and MoS$_2$ (dashed) monolayers layered atop nanopillars of heights $\unit[50]{nm}$ (black), $\unit[100]{nm}$ (green), $\unit[150]{nm}$ (blue) and $\unit[200]{nm}$ (red).}
	    \label{fig:TMD_2d_Topography}
	\end{figure}

A deformation topography of WSe$_{2}$ and MoS$_{2}$ may be drawn (Fig.~\ref{fig:WSe2_3d_Topography}) and compared (Fig.~\ref{fig:TMD_2d_Topography}), deformed by nanopillars within the height range of $\unit[50-200]{nm}$. This range has been chosen to coincide with the experimental possibilities for nanopillar growth and should not strain the monolayers to the point of perforation. These topographies shall lay the foundation of the bandgap renormalisation and conduction band potential calculations performed below.

\section{Strain Induced Potential}
 \label{sec:StrainPot}
 
 With the deformed TMD monolayer topography derived in (\ref{eq:RealSols}), the strain induced potential is given as\cite{pearce2016tight} 
 
  \begin{equation}
    V=\left(\begin{array}{cc}
    \delta_v \mathcal{D} & 0 \\
    0 & \delta_c \mathcal{D} 
    \end{array}\right)
    \label{eq:PotetialMatrixEQ}
 \end{equation}
 
\noindent where $\delta_c$ and $\delta_v$ are the strain response parameters for the conduction and valence bands respectively, and $\mathcal{D}$ is the trace of the strain tensor $\mathcal{D}=\textrm{Tr}[u_{ij}]$. In plate theory, the strain tensor is defined as\cite{landau1986theory}
 
 \begin{figure}[!ht]
     \includegraphics[width=\linewidth]{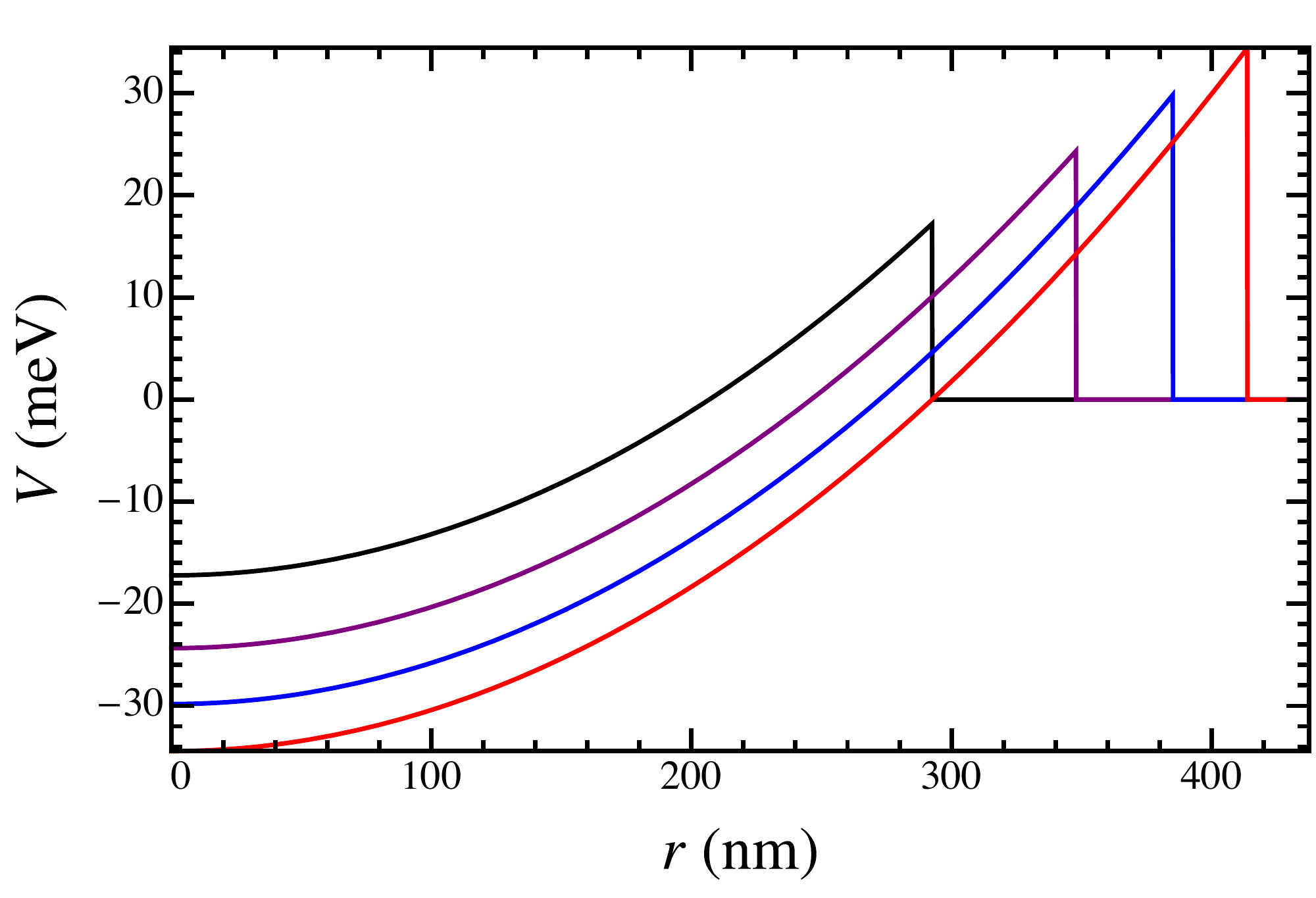}
     \caption{Radial dependance of the strain induced potential in the conduction band of WSe$_{2}$ monolayers deformed by nanopillars of heights $\unit[50]{nm}$ (black), $\unit[100]{nm}$ (green), $\unit[150]{nm}$ (blue) and $\unit[200]{nm}$ (red). The curvature of the potential is unaffected by the pillar height while the depth depends on the height.}
     \label{fig:WSe2_Strain_Potential}
 \end{figure}
 
 \begin{equation}
    u_{ij}=\left(\begin{array}{ccc}
   -h\frac{\partial^2 \zeta}{\partial x^2} & -h\frac{\partial^2\zeta}{\partial x\partial y} & 0 \\
    -h\frac{\partial^2\zeta}{\partial x\partial y} & -h\frac{\partial^2 \zeta}{\partial y^2} & 0 \\
    0 & 0 & \frac{\sigma h}{1-\sigma}\Delta \zeta
    \end{array}\right)
    \label{eq:StrainTensor}
\end{equation}

\noindent Using the above, $\mathcal{D}=\textrm{Tr}[u_{ij}]$ may be simplified to

\begin{equation}
        \mathcal{D}=\frac{(2 \sigma -1 )h}{1-\sigma}\Delta\zeta.
        \label{eq:StrainTensorTrace}
\end{equation}

\noindent Therefore, the strain induced potential in the conduction and valence bands from the derived topography has the following (truncated) harmonic from 
 
\begin{equation}
 V_{c/v}(r)=\begin{cases}
    -\frac{8h\delta_{c/v}(2\sigma-1)(2r^2\beta-\sqrt{H\beta})}{\sigma-1}   & \quad r\leq R\\
   0  & \quad r> R\\
  \end{cases}
  \label{eq:PotetialTMD}
 \end{equation}

\begin{figure}[!t]
     \includegraphics[width=\linewidth]{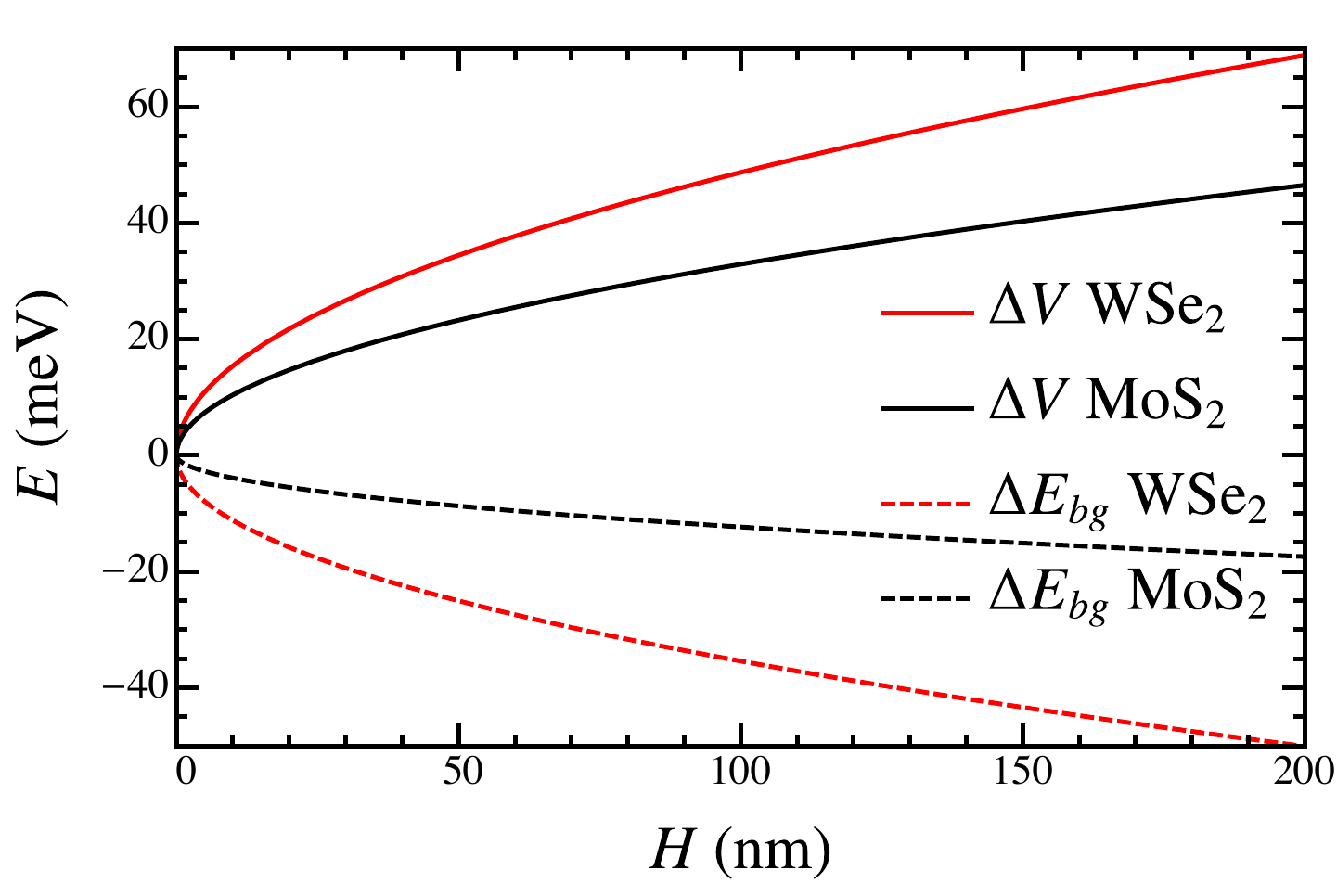}
     \caption{Strain induced potential well height $\Delta V= V(r>R)- V(r=0)$ in the conduction band (solid) and band gap shift $\Delta E_{bg}= E_{bg} -(V_{cb}(r=0)- V_{vb}(r=0)$ (dashed), induced by deforming nanopillars of height $H$ for WSe$_{2}$ monolayers (red) and MoS$_{2}$ monolayers (black).}
     \label{fig:MoS2_Vs_WSe2_Renorms}
 \end{figure}

Notably, as can be seen in Fig.~\ref{fig:WSe2_Strain_Potential}, the height of the deforming nanopillar does not affect the curvature of the induced potentials in the conduction and valence bands, yet does affect the overall bandgap shift (Fig.~\ref{fig:MoS2_Vs_WSe2_Renorms}) and potential well depth (Fig.~\ref{fig:WSe2_Strain_Potential}),  i.e. the difference in potential between $r=0$ and $r>R$. Although our assumptions are modest, this result aligns with experiment\cite{palacios2017large}, where the linewidth of single photon emitters observed at the tip of deforming nanopillars was shown to scale with nanopillar height. In the experiment no descernable relationship between the nanopillar height and the emitted photon energies was observed, most likely due to the uncontrolled topographical variance between the observed strain induced quantum emitters. However, while the shift in photon energy compared to unstrained monlayers seen experimentally are approximately equal to those predicted by the bandgap renormalisation calculated, if the experimental systems offered greater consistency in topography with varying nanopillar height, we predict that there should be a shift in the photon energy by the predicted bandgap shift shown in in Fig.~\ref{fig:MoS2_Vs_WSe2_Renorms}. 

The quadratic form of this potential also allows for further extrapolations of the properties of the strained potential wells to be made such as an estimation of single particle energy spectra and expected leakage.

\section{Fock Darwin Energy Levels}
 \label{sec:FckDrwEn}
 
From DFT studies\cite{chang2013orbital}, the behaviour of the conduction and valence bands under strain for the four most common TMD species (\emph{MX}$_2$ with \emph{M}=Mo, W and \emph{X}=S, Se) is well characterised. Notably, WSe$_2$ is the only compound which exhibits hole attraction, i.e. an increase in the valence band energy about the $K$($K'$) point, under strain. Conversely, the other three TMD species are believed to demonstrate hole repulsion, i.e. an increase in the valence band energy at the $K$($K'$) point, under strain. This is partly why WSe$_2$ has been the material of choice of optical strain experiments searching for quantum emitters in determined strained regions, as the strain potential shape of both bands should allow for exciton funnelling to a strain maximum. It can also be argued that strain induced single particle devices such as quantum dots may be implemented in the other TMD types such as MoS$_2$\cite{feng2012strain}, due to the hole repulsion. 

Here we calculate the single particle energy spectra of the strain induced quantum dots, given deformed topography and induced strain potential described in Sec.~\ref{sec:Deformation} and~\ref{sec:StrainPot}, in the presence of an external magnetic field. We begin by assuming the potential depth of the well described in (\ref{eq:PotetialTMD}) to be deep enough that a harmonic potential may be assumed. Then the Fock-Darwin energy levels of the quadratic portion of the potential may be obtained from the single particle energy given by 7-band $k\cdot p$ theory analysis of an electron in a perpendicular magnetic field $B$ confined in a TMD monolayer\cite{kormanyos2014spin}, combined with the bandgap shift of (\ref{eq:PotetialTMD}). Thus the single band electron energy $E_{n,l}^{\tau,s}$ in a strain-induced potential with external magnetic field is given as

  \begin{equation}
       	E_{n,l}^{\tau,s}=E_{\textrm{FD}}+E_{\textrm{SO}}+E_{\textrm{TRSV}}+E_{\textrm{ZS}}
  	    \label{eq:FckDrwEnLvlsTot}
  \end{equation}
  
\noindent where $\tau=\pm1$ labels the valley isospin $K$($K'$) respectively, $s=\pm1$ labels the electron spin $\uparrow$($\downarrow$) along the $z$ direction respectively. In~\ref{eq:FckDrwEnLvlsTot}, $E_{\textrm{FD}}$ gives the Fock-Darwin energy levels of a 2D harmonic potential quantum dot defined by the strain potential

\begin{equation}
    	\begin{split} E_{\textrm{FD}}=(n+1)\sqrt{\frac{(\hbar\omega^{\tau,s}_c)^2}{4}-\frac{32\hbar^2h\beta\delta_{c}(2\sigma-1)}{m_{\textrm{eff}}^{\tau,s}(\sigma-1)}}&\\
	  -\frac{\hbar\omega^{\tau,s}_c l}{2}+\frac{8h\delta_c(2\sigma-1)\sqrt{H\beta}}{\sigma-1}&
	   \end{split} 
	    \label{eq:FckDrwEnOnly}
\end{equation}

\noindent where $\omega^{\tau,s}_c$ is the cyclotron frequency given by the valley and spin dependant effect mass $m_{\textrm{eff}}^{\tau,s}$. $E_{\textrm{SO}}$ gives the energy splitting due to spin-orbit coupling of the Kramers pairs

\begin{equation}
      	  E_{\textrm{SO}}=\tau s \Delta_{cb} 
		   \label{eq:SOOnly}
\end{equation} 

\noindent where $\Delta_{cb}$ is the splitting in the conduction band about the $K$($K'$) points. $E_{\textrm{TRSV}}$ gives the of the valley degeneracies due to the violations of time-reversal symmetry

\begin{equation}
      	  E_{\textrm{TRSV}}=\frac{(1+\tau)\textrm{sgn}(B)}{2}\hbar\omega^{\tau,s}_c
		   \label{eq:TRSOnly}
\end{equation} 

\noindent and finally $E_{\textrm{ZS}}$ gives the valley and spin Zeeman splitting

 \begin{equation}
       	  E_{\textrm{ZS}}=\frac{\mu_B B}{2}\left(\tau g_{v} +s g_s\right)
 		   \label{eq:ZSOnly}
 \end{equation}
 
\noindent where $g_{v}$ is the valley-Zeeman splitting g-factor and $g_{s}$ is the spin-Zeeman splitting g-factor. The quantum numbers are the principal quantum number $n=0,1,2\dots$, which is defined as $n=2n_r+|l|$ with the radial quantum number $n_r$ of the wavefunction and the azimuthal quantum number $l=-n,-n+2,\dots , n-2,n$. 

\begin{figure}[!b]
    \includegraphics[width=\linewidth]{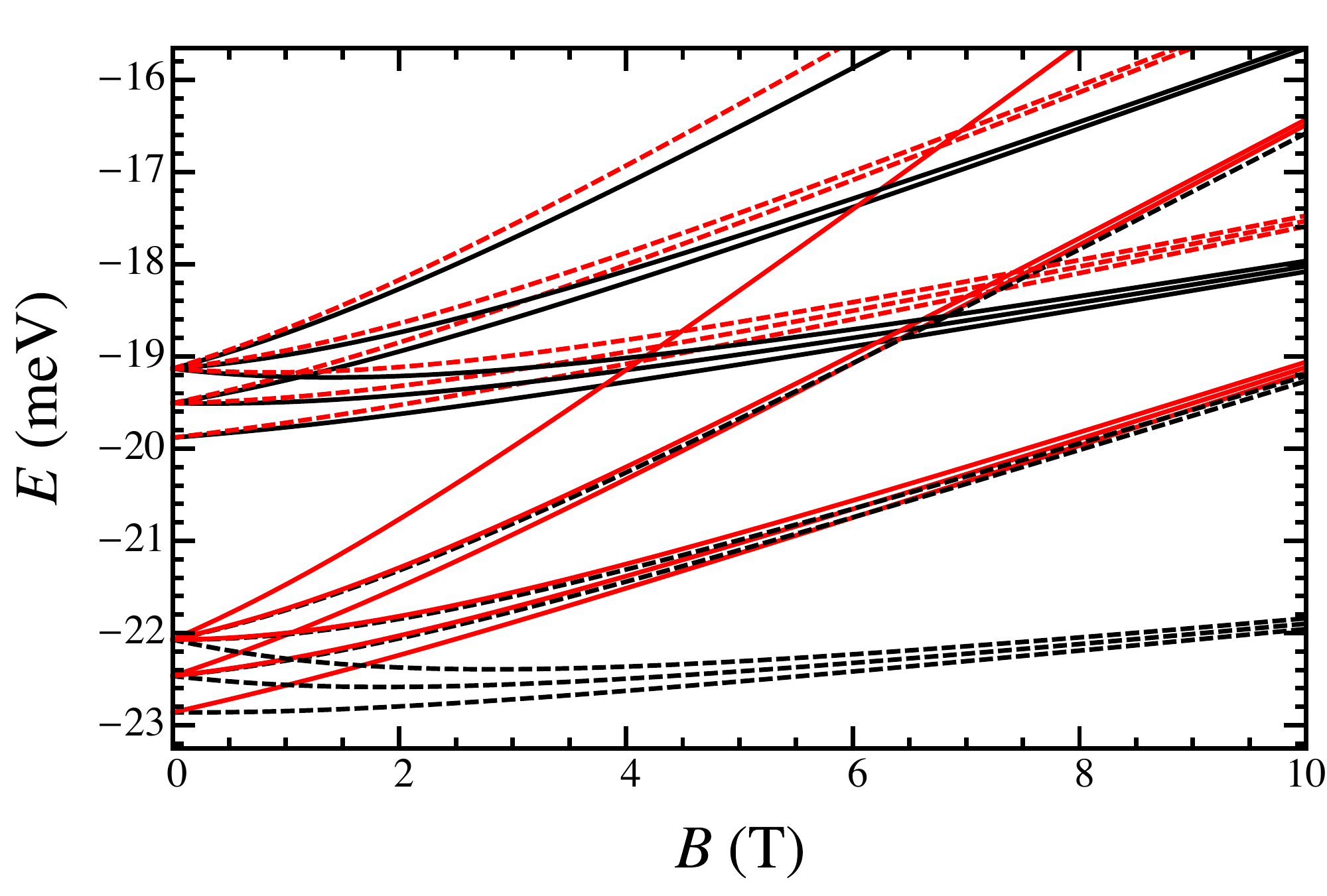}
    \caption{Fock-Darwin energy spectra with external perpendicular magnetic field $B$ of an electron confined within the strain induced potential well of an MoS$_{2}$ monolayer deformed by a nanopillar of height $\unit[200]{nm}$ up to $n=2$ $l=\pm 2,0$. Here, the $K$ $(K')$ states are given by the black (red) lines and the $\uparrow$ $(\downarrow)$ states are given by the solid (dashed) lines.}
    \label{fig:MoS2_FckDrw_Energy_Spectra}
\end{figure} 

The energy spectra with out-of-plane magnetic field of the first few states in MoS$_2$ is shown in Fig.~\ref{fig:MoS2_FckDrw_Energy_Spectra}. The larger spin-orbit splitting and lower magnetic response in WSe$_2$ give rise to a relatively unchanged magnetic spectrum when compared to other confinement methods in TMD monolayers. However, the MoS$_2$ levels demonstrate greater magnetic sensitivity than those derived for quantum dots assuming a hard wall potential of electrostatic gating\cite{kormanyos2014spin}, with clear Landau levels present at magnetic field strengths of $\sim\unit[5]{T}$.

Notably, the energy spectra depicted here are calcuated with constants mimicking previous experimental set-ups\cite{palacios2017large}. As such, the dots assumed are particularly large ($\sim\unit[1]{\mu m}$ in diameter), which limits some of their potential for single particle applications and scalability. In Sec~\ref{sec:Discussion}, possible methods of maintaining the spectra shown here for topographies more conducive to dot applications are discussed. 

\section{State Leakage}
 \label{sec:WKB}
 
One important comparison that needs to be made when comparing strain induced potential wells in TMDs with other confinement methods, is the state leakage probability. Demonstration of low leakage confinement by just out-of-plane straining of the monolayer crystal would open up the discussion for strained TMDs for quantum dots, whose purposes extend past single photon emitters, to single electron dots that may be coupled to other dots in a strain array for quantum information purposes. The transmission of an electron through a potential barrier like the ones discussed in this work may be calculated by the semi-classical WKB method. The unitless transmission coefficient $T$ is given in the following form
 
 \begin{equation}
        T_{n,l}^{\tau,s}=\exp{\left[-\frac{2}{\hbar}\int_{r_0}^{R}\sqrt{2 m_{\textrm{eff}}^{\tau,s}[V(r)-E_{n,l}^{\tau,s}]}dr\right]}
        \label{eq:TransCoef}
\end{equation}

 \begin{figure}[!b]
     \includegraphics[width=\linewidth]{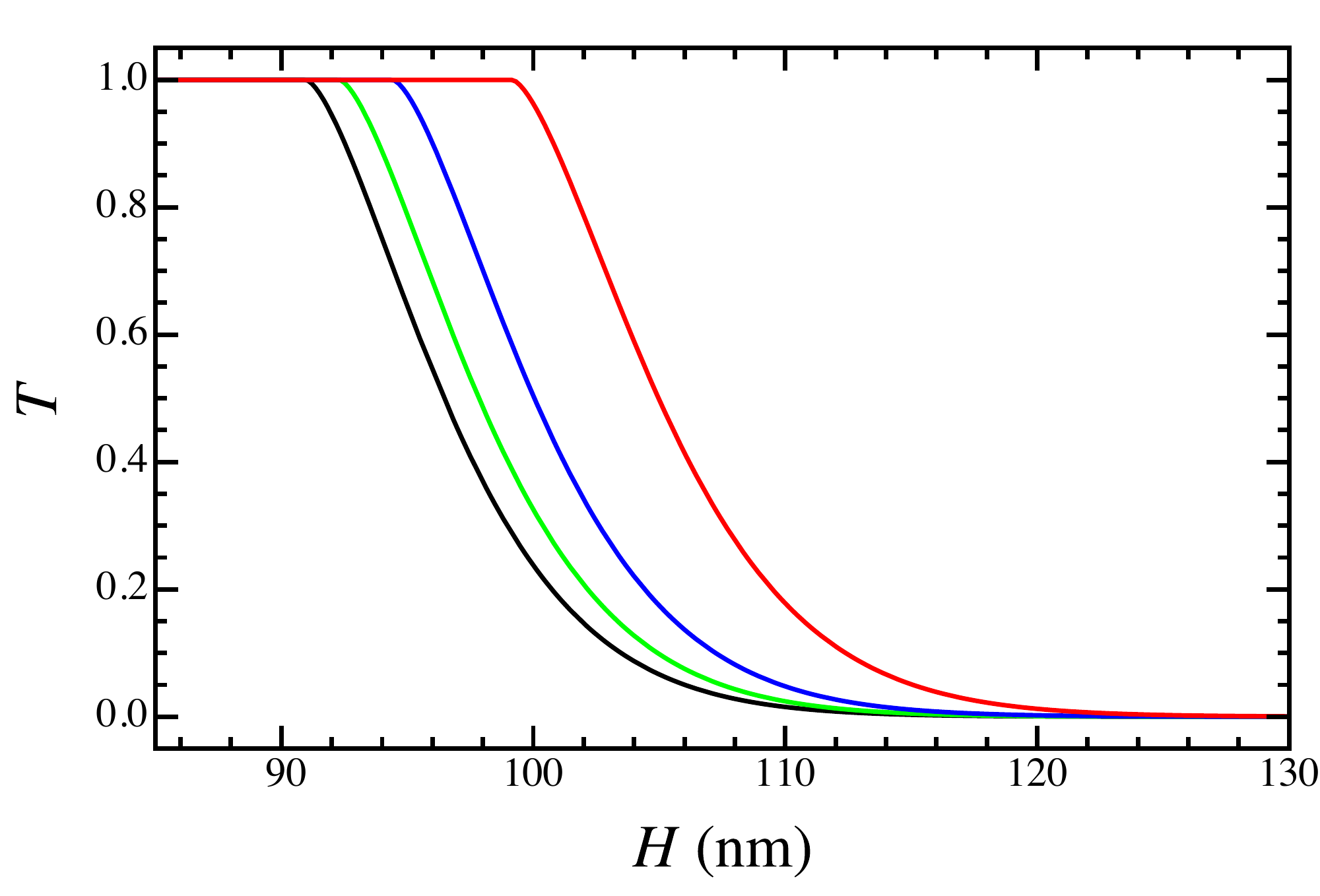}
     \caption{Transmission coefficient of electrons in the $\ket{K\downarrow}$, $n=0$, $l=0$ state with magnetic fields $B=\unit[0]{T}$ (black),  $B=\unit[2.5]{T}$ (green), $B=\unit[5]{T}$ (blue) and $B=\unit[10]{T}$ (red) in potential wells induced by nanopillars of height $H$ in WSe$_{2}$ monolayers as given by the Fock-Darwin energy levels.}
     \label{fig:WSe2_WKB_Transmission}
 \end{figure}

\noindent where $r_0$ is the radial coordinate of the classical turning point at which (\ref{eq:PotetialTMD}) yields $V(r_0)=0$, below which tunneling is not allowed, and above which the WKB approximation is valid. The full form of $r_0$ is given as follows 

 \begin{equation}
        r_0=\sqrt{\frac{\sqrt{H}}{2\beta}+\frac{E_{n,l}^{\tau,s}(\sigma-1)}{16 h \beta \delta_{c}(1-2\sigma)}}.
        \label{eq:r0}
\end{equation}
 
\begin{figure}[!t]
     \includegraphics[width=\linewidth]{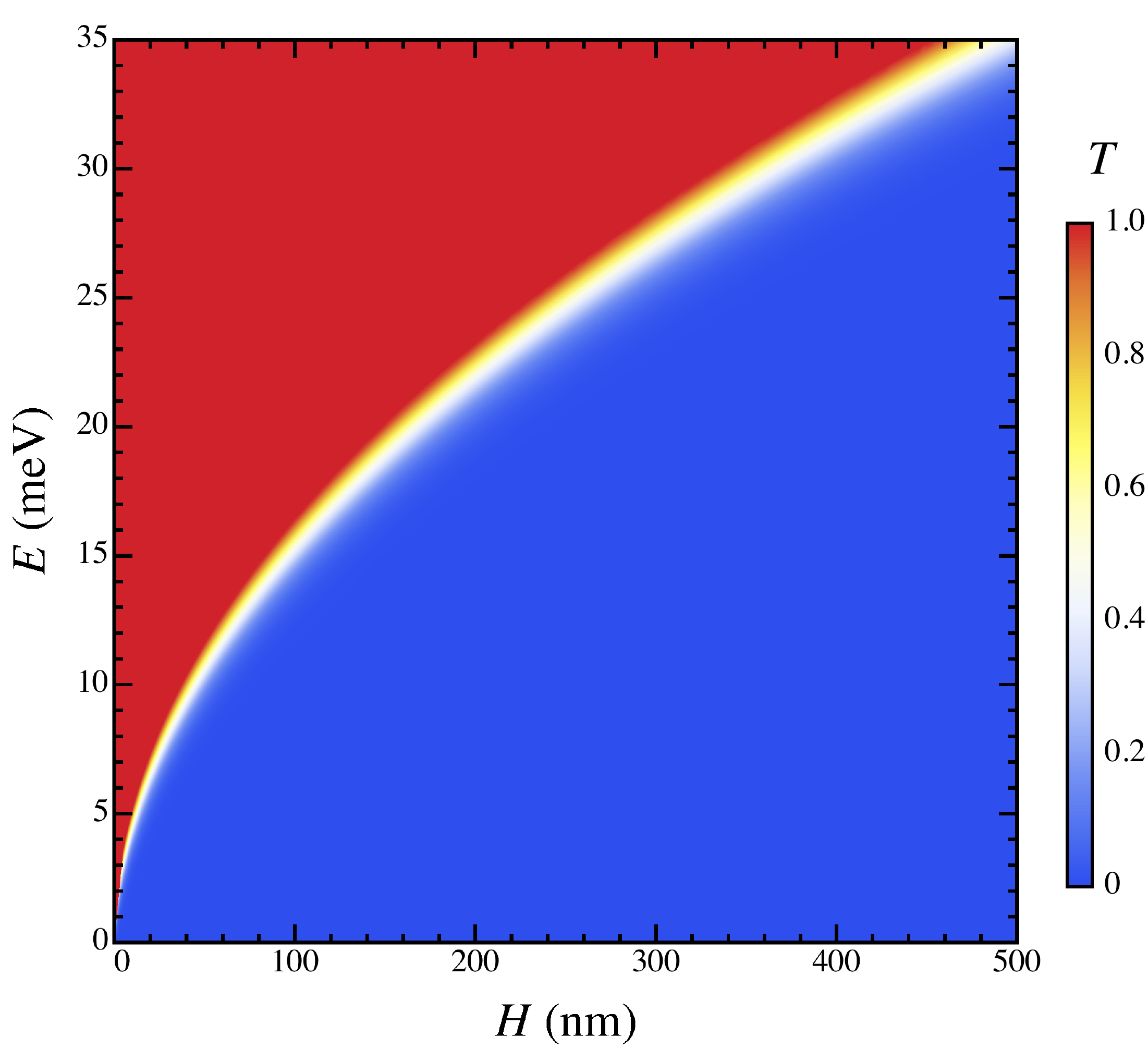}
     \caption{Transmission coefficient spectrum of electrons in MoS$_{2}$ monolayers of energies $E$ in potential wells induced by nanopillars of height $H$.}
     \label{fig:WKB_Transmission_NonSpecific_3D}
 \end{figure}

As can be seen from Figs.~\ref{fig:WSe2_WKB_Transmission} and~\ref{fig:WKB_Transmission_NonSpecific_3D}, the transmission coefficient of electrons out of the potential well in the classically allowed region ($E>0$) is a function of the height of the deforming nanopillar, exhibits a sharp cut-off point at which the electron may be assumed to be well confined.  This sharp dependance of confinement with the nanopillar height aligns well with experimental results demonstrating decreased single photon emitter linewidth with increased nanopillar height\cite{palacios2017large}. As is also visible in Figs.~\ref{fig:WSe2_WKB_Transmission} the WKB approximation breaks down at $T\approx 1$.  


For states below the classically forbidden region ($E<0$) the potential well, no tunnelling outside the strained area should occur, unless aided by some thermal process. For MoS$_2$ this is very promising, as theoretically these strained dot arrays, if prepared properly and held at a sufficiently low temperature, should demonstrate low leakage. Additionally, the potential well depth may be tuned for state selection. If the height of the nanopillar is chosen such that the groundstate energy of one of Kramers pairs lies below the external zero-energy line, while the opposite Kramers pair lies above the line, with time, the dot will deliberately leak the unwanted Kramers pair, isolating only the desired Kramers pair. Additionally, an external magnetic field may be applied perpendicular to the dot to further tune the dot to confine only one spin-valley combination within the low-energy Kramers pair. This state selection process may be difficult in MoS$_2$, since due to the relatively low spin-orbit splitting, high precision in the process may be required. However, TMD monolayers such as W$X_2$, where the spin-orbit splitting is an order of magnitude greater, and hole repulsion due to strain is still present, this Kramers isolation energy alignment may more easily be attained. 
 
This demonstration of possible low leakage confinement in strain dots may be the key to opening up the possibility of strain defined technologies in the TMD monolayers. For example, if an array of confining nanopillars may be grown underneath a TMD monolayer, with patterned electronic gates atop the TMD, the strain would confine the electrons in the dots, while the gates may be used for local potential offsets to tune the coupling between each of the dots.

\section{Discussion}
\label{sec:Discussion}

Many possible electronic, photonic, spintronic and valleytronic applications of TMD monolayers are in discussion, as these materials offer a number of interesting physics. Strain-induced potential engineering is quickly becoming one of the many tools available for device implementation. Strain defined quantum dots and wires in the monolayers can be used in combination with other confinement techniques such as electronic gating and patterned etching or cutting of the monolayers. Additionally, strain engineering is also compatible with Van der Waals few layer heterostructure devices. This extensive toolbox of device engineering may allow for a new dimension of TMD devices to be explored. Here, we discuss the possibility of hybrid device implementations, building on the notion of strained well arrays introduced in Sec.~\ref{sec:WKB}.

The previously introduced idea of an electrically tuneable quantum dot array strained by nanopillares does present some problems, primarily; how would gating such an array affect the exploited mechanical properties of the TMD monolayer? Traditional metal contacts deposited on the TMD will certainly give regions of counterproductive stiffness to the monolayer, either limiting the strain response exploited in this work, or increasing the probability of perforation or rupture. One possible solution is the replacement of the metal contacts with Van der Waals heterostructure contacts. Some realisations of TMD gated devices have used separate graphene sheet contacts either side of the device to tune the potential in lieu of or as intermediate interface with metal contacts\cite{mak2016photonics,bertolazzi2013nonvolatile}. These heterostructure would impact the mechanical properties to a certain extent, stiffening the Young's modulus and Poisson's ratio, but not enough to nullify the results presented here, additionally, the heterostructure should not affect the likelihood of monolayer damage. Positioning will be key in implementing a hybrid heterostructure gated-strain dot array, such as to correctly align nanopillars with the contacts. Thus, thin-finger like graphene nanoribbons or even carbon nanotubes contacts would provide a positioning challenge while reducing the risk from perforation that etching provides. 

\begin{figure}[!b]
    \includegraphics[width=\linewidth]{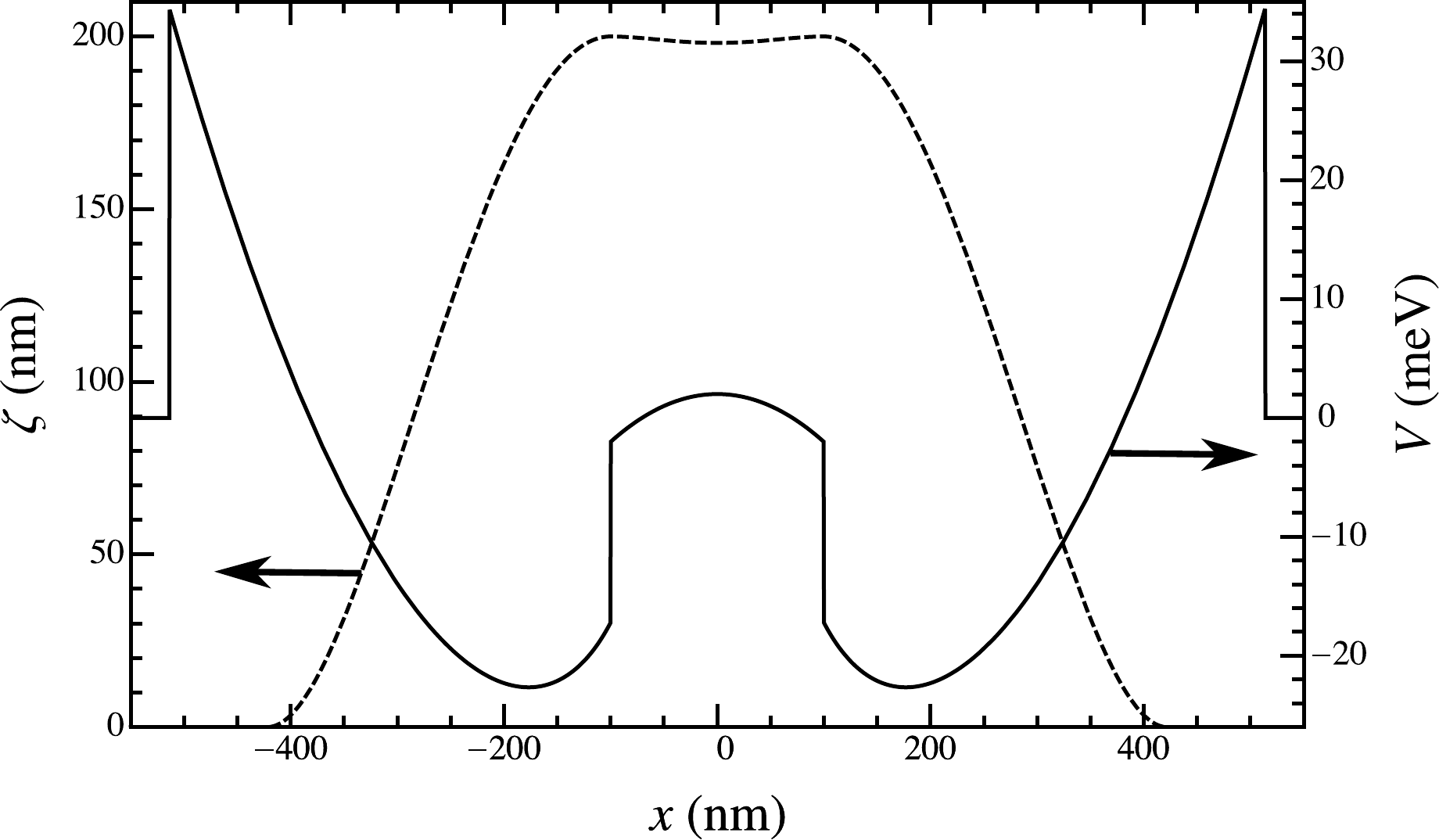}
    \caption{Cross section along the $x$-axis of the height field $\zeta$ (dashed) and potential $V$ (solid) induced in a WSe$_2$ monolayer by two \unit[200]{nm} heigh nanopillars separated \unit[200]{nm} apart centered at $\mathbf{r}=(\pm\unit[100]{nm},0)$.}
    \label{fig:MultiPillar_Pot}
\end{figure}

If two nanopillars are placed close together, with a TMD layered above them, the resulting potential would resemble two anharmonic wells with a near square potential barrier of width equal to that of the distance between the nanopillars (Fig.~\ref{fig:MultiPillar_Pot}). If carbon nanotube contacts are placed atop the TMD in between the nanopillars and on either side of the central strained region, then these contacts may be used to tune the barrier height between the two wells. In WSe$_2$ such a device could have interesting quantum optical applications. As it is known that similar strained regions in WSe$_2$ demonstrate quantum emitters, two dots joined by a tuneable tunnelling barrier would allow for a switchable coupling of the emitted photons from the device. A similar principle could be used with a MoS$_2$ single particle quantum dot, where a variable potential may be used tune the coupling constant of the dots, a necessary feature in scaling semiconducting spin or Kramers qubit implementations. This would allow for control over a quantum dot array while eliminating some of the charge noise compared to a similar system that is purely electrically confined\cite{russ2016coupling}, as fewer metallic gates would be needed to implement such a scheme. 

The work presented has focussed on passively straining monolayer specifically with nanopillars grown from silica substrates. This method has been demonstrated to be useful for exciton coalescing, allowing for arrays of deterministicly placed quantum emitters. For single particle quantum dot applications, the width of such purely strain induced wells may be too wide ($\sim\unit[500]{nm}$) and the well depths too shallow ($\sim\unit[70]{meV}$) to be experimentally useful. However, the method discussed in this work is only straining the monolayers up to $\sim0.5\%$ as calculated from the trace of the strain tensor. This is a very comfortable level of strain for a TMD monolayer, as these materials should be able to withstand straining up to $\sim10\%$ before rupture and $\sim2\%$ before transitioning to an indirect bandgap in $X$S$_2$ type monolayers. More active straining of the TMDs could be implemented to engineer deeper, smaller dots but only up to these material limits. One method of doing so while still using a nanopillar system could be from material selection of the substrate to foster greater Van der Waals attraction between the substrate and the TMD, or by electrostatic attraction of TMD to a backgate underneath the substrate\cite{wong2010characterization}. These methods have good compatibility as the allow for degrees of control over the system parameter $\beta=H/R^4$. AFM tip straining is another proposed method of tuneable active straining for TMD QD definition. This method offers a more addressable height to radius ratio of the dot at a greater risk of perforation, and may not be as compatible with the on chip hybrid systems discussed.

An additional possibility of hybrid implementation is impurity compensation in TMD QD implementations. Experimentally, lattice defect density in TMD monolayers is still problematic, randomly distributing local potential minima within a gated dot region. Similarly to using such strained systems to deterministically place SPEs as opposed of relying on randomly distrbuted lattice defects, an additional layer of strain potential within a gated dot would limit the effect of the random strain defects.

Deterministic straining in combination with more conventional low dimensional device control methods could potentially open up more device possibilities or improved implementations of  devices in TMD monolayers. This further addition to the toolbox of low dimensional material manipulation may help further bolster the already fertile field next generation TMD based technologies. 

\section{Summary}
\label{sec:Summary}

In this work, an analytical description of the deformation topography and strain induced potentials in monolayer TMD over nanopillars is derived from continuum mechanical plate theory. We find a potential well shape that is independant of the pillar height and a sharp drop-off of electron leakage with nanopillar height, as given by a WKB theory analysis, matching observations in experiment while predicting the energy dependance of the emitted photons with nanopillar height. It can then be argued that the resulting strain potentials  from such a setup have further use in hybrid design TMD devices, offering an additional layer of manipulation to a rapidly advancing field of technology. We propose a simple double quantum dot setup using adjacent nanopillars deforming a TMD with conducting graphene heterostructure contacts allowing a tuneable coupling between two dots with fewer gates and thus lower electrical noise than conventional semiconducting quantum dot arrays, and a method of compensating for lattice defects with controlled strain within a traditionally gated TMD QD.

\section{Acknowledgements}
\label{ref:Acknowledgements}

We acknowledge helpful discussions with A. David, A. Korm\'{a}nyos, A. Pearce, M. Russ, V. Shkolnikov and L. Sortino and funding through both the European Union by way of the Marie Curie ITN Spin-Nano and the DFG through SFB 767.

\bibliography{StrainConBibliography}

\end{document}